# The nonextensive parameter for the rotating astrophysical systems with power-law distributions

Yu Haining and Du Jiulin[*]

*Department of Physics, School of Science, Tianjin University, Tianjin 300072, China*

**Abstract** We study the nonextensive parameter for the rotating astrophysical systems with power-law distributions, including both the rotating self-gravitating system and the rotating space plasma. We extend the equation of nonextensive parameter to complex system with arbitrary force field $F(r,v)=F_1(r)+\alpha\, v\times F_2(r)$, and derive a general equation of the $q$-parameter, most generally including both the rotating self-gravitating systems and the rotating space plasmas. At the same time, we reproduce the $\kappa$-distribution in space plasmas and obtain equations of the $\kappa$-parameter. We show that the $q$-parameter is related not only to the temperature gradient, the gravitational force and the electromagnetic force, but also to the inertial centrifugal force and Coriolis force. Thus the rotation introduces significant effect on nonextensivity in the systems. Several examples are given to illustrate the nonextensive effect introduced by the rotation.

-----------------------------------------

In recent years, a lot of attention has been paid to nonextenstive statistical mechanics (NSM), a generalized statistical theory of Boltzmann–Gibbs statistics (BG), so as to deal with the complex systems with power-law distributions. NSM has been proved to be an effective theory for the nonequilibrium complex systems in external fields and therefore has attracted more and more attention. It has been applied to many fields of science and technology in physics, astronomy, chemistry, life science, engineering and even society [1]. Among the astrophysical systems with power-law distributions, self-gravitating systems and space plasmas are the most interesting research areas. For example, in self-gravitating systems, NSM has been applied to study the negative heat capacity [2, 3], the equilibrium structure and stability [4-6], the dark matter distributions [7-9] and the rotations [10-14], etc. In the space plasmas, NSM has been applied to study various waves and instability [15-18] and solar wind [19-21], and properties of the plasmas with the $\kappa$-distributions [22-31]. In particular, we have recently seen detailed studies and deep understanding of the $\kappa$-distributions in the space plasmas and their statistical properties [27-31].

Power-law distributions have been observed frequently in many complex systems including self-gravitating systems and astrophysical plasmas, but the physical meaning of the power-index parameter has still been discussed, particularly, in those complex astrophysical systems having both the gravitational and electromagnetic long-range interactions. In fact, early in 1968 [32], based on using observations of low-energy electrons in the magnetosphere, Vasyliunas analyzed energy spectra of the observed electrons within the plasma sheet and drew an empirical $\kappa$-function to model the velocity distribution of the electrons. This is a power-law distribution and later known as the $\kappa$-distribution. Spacecraft measurements of the plasma velocity distributions both in the solar wind and in the planetary magnetosheaths have revealed that non-Maxwellian distributions are quite common. In many situations the distributions appear reasonably Maxwellian at low energies but have a "suprathermal" power-law tail with the $\kappa$-distribution at high energies. In the solar corona, $\kappa$-like power-law distributions have been proposed to arise from strong nonequilibrium thermodynamic gradients, and temperature anisotropies in the solar wind are correlated with the magnetic fields (see [33] and the references therein). Such astrophysical power-law distributions can be studied partly under the framework of NSM.

---
[*] jldu@tju.edu.cn

In NSM, according to the $q$-kinetic theory, the position-velocity distribution function of the particles in long-range interacting systems can be written by a generalized Maxwell-Boltzmann (MB) distribution (the $q$-distribution),

$$f_q(\bm{r},\bm{v}) = n_q(\bm{r}) B_q \left(\frac{m}{2\pi k_B T}\right)^{3/2} \left[1-(1-q)\frac{m(\bm{v}-\bm{u})^2}{2k_B T}\right]^{1/(1-q)}, \qquad (1)$$

where $n_q$ is the density, $B_q$ is a $q$-dependent normalization constant, $k_B$ is Boltzmann constant, $m$ is particle mass, and $T$ is temperature. For generality, $\bm{u}$ is introduced as the overall bulk velocity of the system. The $q$-parameter which is deviation from one measures the degree of nonextensivity of the system. If we take $q\to 1$, Eq.(1) recovers the Maxwellian distribution. This power-law $q$-distribution could be derived either by the Maxwellian path [34], the nonextensive kinetic theories [35] or by the probabilistically independent postulate based on NSM [36]. In fact, by introducing new fluctuation-dissipation relation [37], the stochastic dynamics with anomalous diffusions can exactly produce the $q$-distributions in a complex system [38,39]. It has been observed that Eq.(1) can excellently model the dark matter haloes that were observed in spherical galaxies with regard to their nonequilibrium stationary states [9]. Usually, the $q$-distribution has been considered equal to the $\kappa$-distributions in the space plasmas [19,22,24-31].

As we know, it is crucial to correctly understand the physical meaning of the nonextensive $q$-parameter in NSM. Here we mainly consider the astrophysical systems and plasmas with long-range interactions. In previous work, the $q$-parameter for the self-gravitating gas was obtained in the equation [40],

$$k_B \nabla T(\bm{r}) + (1-q) m \nabla \varphi_g(\bm{r}) = 0, \qquad (2)$$

where $T(\bm{r})$ is the temperature, $m$ is the particle mass and $\varphi_g(\bm{r})$ is the gravitational potential. For the systems such as galaxies and dark matter haloes [41], $k_B T/m$ is usually replaced by square of the velocity dispersion $\sigma^2$ and then Eq.(2) becomes

$$2\sigma(\bm{r})\nabla\sigma(\bm{r}) = -(1-q)\nabla\varphi_g(\bm{r}). \qquad (3)$$

A similar equation to Eq.(2) of the $q$-parameter was found in the nonequilibrium plasma with Coulomb long-range interactions [42] as

$$k_B \nabla T(\boldsymbol{r}) = e(1-q) \nabla \varphi_c(\boldsymbol{r}). \tag{4}$$

And if it has magnetic field [43], one has

$$k_B \nabla T(\boldsymbol{r}) = e(1-q) [\nabla \varphi_c(\boldsymbol{r}) - c^{-1} \boldsymbol{u} \times \boldsymbol{B}], \tag{5}$$

where $e$ is the electron charge, $\varphi_c(\boldsymbol{r})$ is the Coulomb potential, $c$ is the light speed, $\boldsymbol{B}$ is the magnetic induction intensity. Thus the equation of the $q$-parameter contains the contribution not only from the Coulomb potential, but also from the magnetic field in the plasma. These equations tell us a fact that the power-law $q$-distribution for $q \neq 1$ can represent a nonequilibrium stationary distribution of the systems in external force fields. When $q=1$, the systems are in an isothermal equilibrium state. Eq.(2) has been applied to test the nonextensivity in the helioseismological measurements of the Sun [44], to constitute the astrophysical convection instability criterion [45,46], to analyze the thermodynamic stability of self-gravitating system [47], and to introduce the concept of the gravitational heat conduction [48] and the gravitational temperature [49]. Eq.(4) has also be applied to establish new characteristics in the nonequlibrium space plasmas with the power-law distributions [17,18,50,51].

In this work, on the basis of the $q$-kinetic theory in NSM we will study the $q$-parameter for the rotating astrophysical systems with the $q$-distribution, generally including the self-gravitating systems and the space plasmas. We notice that so far the theories for the $q$-parameter have not considered the rotation effect of astrophysical systems. Rotation is the most common phenomenon in the astrophysical systems such as planets, stars, galaxies and the space plasmas, for example, the fixed-axis rotation with uniform angular velocity. In the self-gravitating systems, the rotation effect on the equilibrium property in NSM has recently attracted attentions [10-14]. Due to high-speed revolution, the rotational symmetry can be spontaneously broken down and as a result the non-trivial density profiles corresponding to complex equilibrium configurations can be generated [52]. In addition, it is known that the obvious shift to the east or to the west of the trade winds of the Earth's atmosphere is caused generally by the rotation effect.

The paper is organized as follows. Firstly we extend the study of the $q$-parameter to the system under a very general force field so as to obtain a general equation of the $q$-parameter in complex systems with power-law distributions. And then we apply it to the rotating self-gravitating system and the rotating space plasma, respectively, to derive the equations of the $q$-parameter in these systems. We give discussion of the rotation effect on the nonextensivity. We take the Sun, Jupiter and Saturn as examples to illustrate the nonextensive effect introduced by the rotation. Finally, we present the conclusions.

We consider a self-gravitating system rotating about a fixed-axis with uniform angular velocity and we analyze it in a space rotation reference frame. Then the force field on the particle can be expressed by the space/velocity-dependent force,

$$\boldsymbol{F}(\boldsymbol{r}, \boldsymbol{v}) = -m\nabla \varphi_g(\boldsymbol{r}) + m\omega^2 \boldsymbol{R} + 2m\boldsymbol{v} \times \boldsymbol{\omega}, \tag{6}$$

where the first term is the gravitational force in inertial frame, the second is the inertial centrifugal force and the third corresponds to the Coriolis force. The quantity $\omega^2 \boldsymbol{R}$ is often called centrifugal acceleration, $\boldsymbol{R}$ is the vertical distance between the particle and the rotation axis and its orientation is outward and perpendicular to the axis, $\boldsymbol{\omega}$ is the rotation angular velocity, and $\boldsymbol{v}$ is the velocity of the particle.

Now we consider the rotating space plasma with the same physical situation as above, then the force field on the particle can be expressed by the space/velocity -dependent force as

$$\boldsymbol{F}(\boldsymbol{r},\boldsymbol{v}) = e\nabla\varphi_c(\boldsymbol{r}) - \frac{e}{c}(\boldsymbol{v}\times\boldsymbol{B}) + m\omega^2\boldsymbol{R} + 2m\boldsymbol{v}\times\boldsymbol{\omega}, \tag{7}$$

where the first two terms are the electromagnetic interaction forces in inertial frame, the third term is the inertial centrifugal force and the fourth corresponds to the Coriolis force, which come from the rotation of the plasma.

In order to study more generally complex systems with power-law distribution, we can write the forces, Eq.(6) and Eq.(7), in a unified form as

$$\boldsymbol{F}(\boldsymbol{r},\boldsymbol{v}) = \boldsymbol{F}_1(\boldsymbol{r}) + \alpha\boldsymbol{v}\times\boldsymbol{F}_2(\boldsymbol{r}), \tag{8}$$

where $\boldsymbol{F}_1(\boldsymbol{r})$ is the force by a potential field, $\boldsymbol{F}_2(\boldsymbol{r})$ is a vector and $\alpha$ is a scalar. Eq.(8) becomes an arbitrary force. For example, if the system is a rotating self-gravitating system, we have that $\boldsymbol{F}_1(\boldsymbol{r}) = -m\nabla\varphi_g(\boldsymbol{r}) + m\omega^2\boldsymbol{R}$ (the sum of the gravitational force and the centrifugal force), the scalar $\alpha = 2m$ and the vector $\boldsymbol{F}_2(\boldsymbol{r}) = \boldsymbol{\omega}$ is the angular velocity. If the system is the rotating plasma with a magnetic field, we have that $\boldsymbol{F}_1(\boldsymbol{r}) = e\nabla\varphi_c(\boldsymbol{r}) + m\omega^2\boldsymbol{R}$ (the sum of the Coulomb force and the centrifugal force), the scalar $\alpha = -(e/c)+2m$, and the vector $\boldsymbol{F}_2(\boldsymbol{r}) = \boldsymbol{B}+\boldsymbol{\omega}$ (the sum of the magnetic induction and the angular velocity).

Now we study the $q$-parameter in the nonextensive system acted upon by the arbitrary force field in the form of Eq.(8). If we use $f(\boldsymbol{r},\boldsymbol{v},t)$ to be the time-dependent space/velocity distribution function of the particles, it is well known that the dynamics of the system under the force field Eq.(8) is governed by the following Boltzmann equation [53],

$$\frac{\partial f}{\partial t} + \boldsymbol{v}\cdot\frac{\partial f}{\partial \boldsymbol{r}} + \frac{1}{m}[\boldsymbol{F}_1(\boldsymbol{r}) + \alpha\boldsymbol{v}\times\boldsymbol{F}_2(\boldsymbol{r})]\cdot\frac{\partial f}{\partial \boldsymbol{v}} = C(f), \tag{9}$$

In NSM, the nonextensive effects can be incorporated through the $q$-collision term $C_q$ in the $q$-kinetic theory, then Eq.(9) becomes the generalized Boltzmann equation [35],

$$\frac{\partial f_q}{\partial t} + \boldsymbol{v}\cdot\frac{\partial f_q}{\partial \boldsymbol{r}} + \frac{1}{m}[\boldsymbol{F}_1(\boldsymbol{r}) + \alpha\boldsymbol{v}\times\boldsymbol{F}_2(\boldsymbol{r})]\cdot\frac{\partial f_q}{\partial \boldsymbol{v}} = C_q(f_q). \tag{10}$$

It was proved that the solutions of Eq.(10) satisfied the generalized $H$-theorem (a $H$-theorem based on the $q$-entropy, see [35]) and would evolve irreversibly towards the $q$-distribution in Eq.(1), the generalized MB distribution. At this time, the system arrives at the $q$-equilibrium and the $q$-collision term on the right hand side of Eq.(10) vanishes. Through $C_q(f_q)=0$, the power-law $q$-distribution Eq.(1) can be obtained.

We can write the $q$-distribution Eq.(1) [36, 41] as

$$f_q^{1-q} = \left(n_0 A_q T^{-3/2}\right)^{1-q}\left[1-(1-q)\frac{m\varphi_g}{k_B T}\right]\left[1-(1-q)\frac{m(\boldsymbol{v}-\boldsymbol{u})^2}{2k_B T}\right], \tag{11}$$

with $A_q \equiv (m/2k_B)^{3/2} B_q$. Such a Hamiltonian factorization was established under the probabilistically independent postulate [36], i.e., under the assumption that the kinetic degrees of freedom are independent (uncorrelated) with the potential degrees of freedom [29-31]. Following the line of Ref.[41], we can study the equation of the $q$-parameter for the nonextensive system with the general force field Eq.(8). At the $q$-equilibrium, Eq.(10) becomes

$$\mathbf{v}\cdot\nabla f_q + \frac{1}{m}[\mathbf{F}_1(\mathbf{r}) + \alpha\mathbf{v}\times\mathbf{F}_2(\mathbf{r})]\cdot\nabla_v f_q = 0, \quad (12)$$

where we have used the symbols $\nabla = \partial/\partial\mathbf{r}$ and $\nabla_v = \partial/\partial\mathbf{v}$. Furthermore, Eq.(12) can also be written in another form,

$$\mathbf{v}\cdot\nabla f_q^{1-q} + \frac{1}{m}[\mathbf{F}_1(\mathbf{r}) + \alpha\mathbf{v}\times\mathbf{F}_2(\mathbf{r})]\cdot\nabla_v f_q^{1-q} = 0. \quad (13)$$

From Eq.(11) we can find that

$$\nabla f_q^{1-q} = -\frac{3}{2}D_q\frac{\nabla T}{T}\left[1-(1-q)\frac{m\varphi_g}{k_B T}\right]\left[1-(1-q)\frac{m(\mathbf{v}-\mathbf{u})^2}{2k_B T}\right]$$
$$-D_q\frac{m}{k_B}\left[1-(1-q)\frac{m(\mathbf{v}-\mathbf{u})^2}{2k_B T}\right]\nabla\left(\frac{\varphi_g}{T}\right) \quad (14)$$
$$+D_q\left[1-(1-q)\frac{m\varphi_g}{k_B T}\right]\frac{m(\mathbf{v}-\mathbf{u})^2}{2k_B T^2}\nabla T,$$

and

$$\nabla_v f_q^{1-q} = -D_q\left[1-(1-q)\frac{m\varphi_g}{k_B T}\right]\frac{m(\mathbf{v}-\mathbf{u})}{k_B T}, \quad (15)$$

where we have used $D_q \equiv (n_0 A_q T^{-3/2})^{1-q}(1-q)$. Substituting Eqs.(14) and (15) into Eq.(13) and using the vector formula, $(\mathbf{a}\times\mathbf{b})\cdot\mathbf{c} = (\mathbf{c}\times\mathbf{a})\cdot\mathbf{b} = (\mathbf{b}\times\mathbf{c})\cdot\mathbf{a}$, we can derive the equation,

$$-\frac{3}{2}\frac{\nabla T}{T}\left[1-(1-q)\frac{m\varphi_g}{k_B T}\right]\left[1-(1-q)\frac{m(\mathbf{v}-\mathbf{u})^2}{2k_B T}\right]$$
$$-\frac{m}{k_B}\left[1-(1-q)\frac{m(\mathbf{v}-\mathbf{u})^2}{2k_B T}\right]\nabla\left(\frac{\varphi_g}{T}\right) \quad (16)$$
$$+\left[1-(1-q)\frac{m\varphi_g}{k_B T}\right]\frac{m(\mathbf{v}-\mathbf{u})^2}{2k_B T^2}\nabla T - \left[1-(1-q)\frac{m\varphi_g}{k_B T}\right]\left[\frac{\mathbf{F}_1+\alpha\mathbf{u}\times\mathbf{F}_2}{k_B T}\right] = 0.$$

In Eq.(16), the sum of the coefficients of the zero power and the second power for the velocity $\mathbf{v}$ must be zero, respectively, because $\mathbf{r}$ and $\mathbf{v}$ are independent variables. Thus, for the coefficient equation of the terms of $(\mathbf{v}\text{-}\mathbf{u})^0$, we obtain

$$\frac{3}{2}\frac{\nabla T}{T}\left[1-(1-q)\frac{m\varphi_g}{k_B T}\right] + \frac{m}{k_B}\nabla\left(\frac{\varphi_g}{T}\right) + \frac{\mathbf{F}_1+\alpha\mathbf{u}\times\mathbf{F}_2}{k_B T}\left[1-(1-q)\frac{m\varphi_g}{k_B T}\right] = 0. \quad (17)$$

For the coefficient equation of the terms of $(\mathbf{v}\text{-}\mathbf{u})^2$, we obtain

$$\frac{3}{2}(1-q)\frac{\nabla T}{T}\left[1-(1-q)\frac{m\varphi_g}{k_B T}\right] + (1-q)\frac{m}{k_B}\nabla\left(\frac{\varphi_g}{T}\right) + \left[1-(1-q)\frac{m\varphi_g}{k_B T}\right]\frac{\nabla T}{T} = 0. \quad (18)$$

Combining Eq.(17) with Eq.(18), we can find the general equation of the $q$-parameter,

$$k_B \nabla T = (1-q)\left[ \boldsymbol{F}_1(\boldsymbol{r}) + \alpha \boldsymbol{u} \times \boldsymbol{F}_2(\boldsymbol{r}) \right]. \tag{19}$$

This is a general form of the equation of the *q*-parameter for the system affected by the arbitrary force field Eq.(8). Eq.(19) shows us clearly that $q \neq 1$ holds if and only if the temperature gradient is $\nabla T \neq 0$ (the temperature is spatially inhomogeneous). In this condition the system is nonextensive and the *q*-distribution represents a stationary nonequilibrium distribution in the complex system. When *q*=1, the temperature is spatially uniform and so the system is in a thermal equilibrium state and is extensive. Thus nonextensivity can be explained as a non-additive property of the nonequlibrium complex system affected by external force fields. The degree of nonextensivity (the *q*-parameter deviation from one) depends on the competition between the thermodynamic force $k_B \nabla T$ and the external force fields $\boldsymbol{F}_1(\boldsymbol{r}) + \alpha \boldsymbol{v} \times \boldsymbol{F}_2(\boldsymbol{r})$ in Eq.(19). If *q*<1, the two forces are the same direction and then the system is super-extensive; if *q*>1, the two forces are reverse and then the system is sub-extensive.

Now we use the *q*-distribution (1) to obtain the mean kinetic energy, and so define the physical temperature $\tilde{T}$ in a nonextensive system [27,30], that

$$\left\langle \frac{1}{2}mv^2 \right\rangle = \frac{3k_B T}{7-5q} \equiv \frac{3}{2}k_B \tilde{T}, \quad 0 < q < \frac{7}{5}. \tag{20}$$

And then, making the replacements in (1), $T = \frac{1}{2}(7-5q)\tilde{T}$ and $(q-1)^{-1} = \kappa + 1$, we can exactly reproduce the *κ*-distribution in the astrophysical and space plasmas,

$$f_\kappa(\boldsymbol{r},\boldsymbol{v}) \sim \left[ 1 + \frac{1}{\kappa - \frac{3}{2}} \frac{m(\boldsymbol{v}-\boldsymbol{u})^2}{2k_B \tilde{T}} \right]^{-(\kappa+1)}, \quad \kappa > \frac{3}{2}. \tag{21}$$

Correspondingly, (19) is written with the *κ*-parameter and the physical temperature by

$$\left( \kappa - \frac{3}{2} \right) k_B \nabla \tilde{T} = -\left[ \boldsymbol{F}_1(\boldsymbol{r}) + \alpha \boldsymbol{u} \times \boldsymbol{F}_2(\boldsymbol{r}) \right]. \tag{22}$$

This is an equation of the *κ*-parameter, which gives the physical meaning that the plasma with *κ*-distribution is in the state to thermal equilibrium if and only if the *κ*-parameter is in the limit to infinity.

On the basis of various physical conditions, Eq.(19) can be applied to both the rotating self-gravitating systems and the rotating space plasmas to obtain their expressions of the *q*-parameter.

***The rotating self-gravitating systems*** -- For the rotating self-gravitating system, based on Eq.(8) we apply the force Eq.(6) to Eq.(19), the equation of the *q*-parameter becomes

$$k_B \nabla T(\boldsymbol{r}) = m(1-q)\left[ -\nabla \varphi_g(\boldsymbol{r}) + \omega^2 \boldsymbol{R} + 2\boldsymbol{u} \times \boldsymbol{\omega} \right]. \tag{23}$$

In the square bracket on the right side, the first term represents contribution of the gravitation to the nonextensivity, the second and third terms are contribution of the rotation to the nonextensivity. It is shown that the nonextensivity is related not only to the temperature gradient and the gravitational force, but also to the inertial centrifugal acceleration, the rotation angular velocity $\boldsymbol{\omega}$ and the overall bulk velocity of the system. So Eq.(23) extends the original equation of the *q*-parameter in Eq.(2) to the rotating system where the stellar differential rotation [54] plays an important role. Eq.(23)

contains the contributions of the inertial centrifugal force and the Coriolis force to the nonextensivity, and therefore extends Eq.(2) to the rotating self-gravitating system.

In this case, the equation of the $\kappa$-parameter, (22), becomes

$$(\kappa - \frac{3}{2})k_B \nabla \tilde{T}(r) = m\,[\nabla \varphi_g(r) - \omega^2 R - 2u \times \omega]. \tag{24}$$

Usually, if the self-gravitating systems are galaxies and dark matter haloes, $k_B T/m$ is replaced by square of the velocity dispersion $\sigma^2(r)$, then the $q$-distribution function is written (see [41]) as

$$f_q(v,r) = n_q(r) B_q \left(2\pi\sigma^2\right)^{-\frac{3}{2}} \left[1 - (1-q)\frac{(v-u)^2}{2\sigma^2}\right]^{1/(1-q)}, \tag{25}$$

where the density distribution function is

$$n_q(r) = n_0 \left[1 - (1-q)\frac{\varphi_g(r)}{\sigma^2}\right]^{1/(1-q)}. \tag{26}$$

In this situation, the equation of the $q$-parameter Eq.(23) can be written as

$$2\sigma(r)\nabla\sigma(r) = (1-q)\left[-\nabla\varphi_g(r) + \omega^2 R + 2u \times \omega\right]. \tag{27}$$

It is clear that we have generalized the $q$-parameter equation Eq.(3) to the rotating self-gravitating system. The nonextensivity exists (i.e. the $q$-parameter deviation from one) if and only if there is the inhomogeneous velocity dispersion (i.e. $\nabla\sigma(r)\neq 0$) in the system. If the velocity dispersion is spatially homogeneous, $\nabla\sigma(r)=0$, then we have $q=1$ and so the system is extensive.

***The rotating space plasmas*** -- For the rotating space plasma with magnetic field, based on Eq.(8) we apply the force Eq.(7) to Eq. (19), and then obtain the equation of the $q$-parameter,

$$k_B \nabla T(r) = (1-q)\,[e(\nabla\varphi_c(r) - c^{-1} u \times B) + m(\omega^2 R + 2u \times \omega)]. \tag{28}$$

In a nonequlibrium complex plasma, on the right side of Eq.(28), the term $e(\nabla\varphi_c(r) - c^{-1} u \times B)$ is the contribution of the electromagnetic fields to the nonextensivity, and the term $m(\omega^2 R + 2u \times \omega)$ is the contribution of the rotation to the nonextensivity. If the plasma is in thermal equilibrium state, we have $\nabla T=0$ and $q=1$, and then the plasma is extensive. Thus the $q$-parameter in the space plasma is related not only to the temperature gradient and the electromagnetic fields, but also to the angle velocity and the overall bulk velocity of the rotating plasma, where the rotation effect plays an important role. It is clear that Eq.(28) can be exactly reduced to original forms in Eqs.(4) and (5) if the rotation effect and/or the magnetic field may be neglected. Therefore we have generalized the equation of the $q$-parameter to the rotating space plasma with power-law distribution.

In this case, the equation of the $\kappa$-parameter, (22), is rewritten as

$$(\kappa - \frac{3}{2})k_B \nabla \tilde{T}(r) = e[-\nabla\varphi_c(r) + c^{-1} u \times B] - m(\omega^2 R + 2u \times \omega). \tag{29}$$

Now we focus on analysis of the nonextensive effect introduced by the rotation. In the above equations (23) and (27)-(29), the contribution of the rotation to the degree of nonextensivity of the nonequilibrium astrophysical systems (self-gravitating system and space plasma) contains two terms, $m\omega^2 R + 2m u \times \omega$. The first term $m\omega^2 R$ is due to the inertial centrifugal force. When the angle velocity $\omega$ varies from the equator to poles, the differential rotation [54] exists depending on the angle velocity $\omega$ and the vertical distance $R$ between the particle and the rotation axis. The second term $2m\mathbf{u}\times\boldsymbol{\omega}$ is due to the Coriolis force, depending on the angle velocity $\omega$ and the convective motion velocity $\mathbf{u}$ of the rotating fluid. Obviously, stellar differential rotation and its combination with convection both produce a change in the degree of nonextensivity in the rotating astrophysical systems, and thereby lead to a change in the power-law space-velocity distribution of the interacting particles. However, the equations of $q$-parameter have no direct connection with stellar rotation period [55] and deviation from spherical symmetry.

Of course, this result may be suitable for understanding the nonextensivity and power-law space-velocity distribution of rotating galaxies and dark matter. In the galaxy and dark matter halo having spatially inhomogeneous velocity dispersion, the differential rotation may introduce differential distribution.

It is usually not easy to study the physical source of the nonextensivity and the power-law distributions in a complex system. This work will help us to understand the nonextensivity introduced by the rotation in astrophysical systems. The reader may see [37,38] about the dynamical origin of the power-law distributions in a complex Brownian motion system, which show that the power-law distributions is due to a new fluctuation-dissipation relation between the diffusion and friction coefficients depending on the energy.

**Examples** -- Eq.(23) shows that the role of the rotation in the nonextensivity not only depends on the angular velocity $\omega$, but also is proportional to the vertical distance $R$ away from the rotation axis, and therefore is the maximum at the equator. In other words, the nonextensive effect introduced by the rotation of the self-gravitating system can be determined by the centrifugal acceleration.

If a rotating self-gravitating system is approximately regarded as spherical symmetry, let $Q=1-q$, Eq.(23) can be written as one-dimension form (here we have assumed the overall bulk velocity $\mathbf{u}=0$),

$$Q = -\frac{k_B}{m}\frac{dT}{dr} \bigg/ \left(\frac{d\varphi_g}{dr} - \omega^2 R\right), \qquad (30)$$

where $-d\varphi_g/dr$ and $\omega^2 R$ are the gravitational acceleration and the centrifugal acceleration, respectively. We use $Q_0$ to denote the nonextensive parameter $Q$ without the rotational effect in Eq. (2), *i.e.*,

$$Q_0 = -\frac{k_B}{m}\frac{dT}{dr} \bigg/ \frac{d\varphi_g}{dr}. \qquad (31)$$

From Eq.(30) and Eq.(31) we get a relative nonextensive parameter introduced by the rotation,

$$\Delta Q = \frac{Q - Q_0}{Q_0} = -\frac{\omega^2 R}{(d\varphi_g/dr) - \omega^2 R}. \qquad (32)$$

Now we take three examples (the Sun, Jupiter and Saturn) in the solar system to illustrate the degree of nonextensivity introduced by the rotation at the equator surface. The relevant data are gotten from Ref.[56]. Table 1 listed the centrifugal acceleration, the gravitational acceleration and the values of $\Delta Q$ at the equators. It is shown that the relative nonextensive parameter introduced by the rotations at the surfaces of the Sun is negligible, so the previous works on the solar without considering the rotation are still highly reliable. However, at the surfaces of Jupiter and Saturn, the nonextensive effects introduced by the rotation are obvious; the relative nonextensive parameters are reduced by 9.7% and 20.5%, respectively.

Table 1: The values of the Sun, Jupiter and Saturn at the equatorial surfaces

| Examples | $\omega^2 R$ (m/s$^2$) | $-d\varphi_g/dr$ (m/s$^2$) | $\Delta Q$ |
|---|---|---|---|
| The Sun | $6.0\times10^{-3}$ | -274.0 | $-2.2\times10^{-3}$% |
| Jupiter | 2.2 | -24.9 | -9.7% |
| Saturn | 1.8 | -10.6 | -20.5% |

In conclusion, we have studied the equation of the nonextensive parameter for the rotating astrophysical systems. The whole analyses in the context are based on the standard kinetic theory in NSM. For the generality, we have studied the equation of the $q$-parameter for the nonequilibrium systems in an arbitrary force field, having a general form of Eq.(8), $\boldsymbol{F}(\boldsymbol{r},\boldsymbol{v})=\boldsymbol{F}_1(\boldsymbol{r})+\alpha\,\boldsymbol{v}\times\boldsymbol{F}_2(\boldsymbol{r})$. We have obtained the equation of the $q$-parameter in Eq.(19) and extended it to the nonextensive complex system acted upon by the most general force above. At the same time, we have reproduced the $\kappa$-distribution and obtained an equation of the $\kappa$-parameter for the space plasmas.

When we apply the $q$-parameter Eq.(19) to the rotating self-gravitating systems with the force in Eq.(6), and we have extended the equation of the $q$-parameter to more general forms, Eq.(23) and Eq.(27) for the two types of the self-gravitating systems, which contains the contributions from the inertial centrifugal force and the Coriolis force to the nonextensivity. It has been shown that the $q$-parameter not only depends on the temperature gradient (thermodynamic force) and the gravitational force, but also depends on the inertial centrifugal acceleration $\omega^2 \boldsymbol{R}$ (the angular velocity $\boldsymbol{\omega}$, and the vertical distance $\boldsymbol{R}$ away from the rotation axis) and the Coriolis acceleration $2\boldsymbol{u}\times\boldsymbol{\omega}$ (the overall bulk velocity $\boldsymbol{u}$ of the rotating fluid), so that stellar differential rotation and its combination with convection can both produce a change in the degree of nonextensivity and thereby in the power-law distribution. Thus the rotation can introduce new nonextensive effect in the systems. For the rotating galaxies and dark matter haloes, the $q$-parameter equation, Eq. (27), has exactly associated the rotation effect (the inertial centrifugal force and the Coriolis force) with the degree of nonextensivity in the self-gravitating systems with spatially inhomogeneous velocity dispersion, which can help to understand the evolution in the space-velocity distributions of the galaxies and dark matter haloes.

When we apply the $q$-parameter Eq.(19) to the rotating space plasma acted on by the force given in Eq.(7), we have also extended the equation to a more general form Eq.(28) containing the inertial centrifugal force and the Coriolis force, and so obtained the rotation effect on the $q$-parameter. It is shown that the $q$-parameter not only depends on the temperature gradient and the electromagnetic fields, but also depends on the inertial centrifugal force and the Coriolis force. The differential rotation and its combination with convection in the space plasma both produce a change in the degree of nonextensivity and thereby have influence on the power-law distribution in the plasma. The equation of the *κ*-parameter is also given in (29).

In addition, we have defined a relative nonextensive parameter in Eq.(32) for the rotating self-gravitating system and took the Sun, Jupiter and Saturn as examples to illustrate the nonextensive effect introduced by the rotation at the equator surfaces. The rotation introduces significant effects on the nonextensive parameter in the systems with power-law distributions.

* * *

The authors thank Dr. G. Livadiotis for his helpful discussions on the kappa-distributions, the physical temperature and the nonextensive statistics in space plasmas. This work is supported by the National Natural Science Foundation of China under grants No 11175128 and 10675088.

REFERENCES

[1] TSALLIS C., *Introduction to Nonextensive Statistical Mechanics: Approaching a Complex World,* (Springer, New York) 2009.
[2] SILVA R. and ALCANIZ J. S., *Phys. Lett. A*, **313** (2003) 393; *Physica A,* **341** (2004) 208.
[3] LIU L. Y., LIU Z. P. and GUO L. N., *Physica A,* **387** (2008) 5768.
[4] LIMA J. A. S., SILVA R. and SANTOS J., *Astron. Astrophys.,* **396** (2002) 309.
[5] DU J. L., *Phys. Lett. A,* **320** (2004) 347; *Physica A,* **335** (2004) 107.
[6] DU J. L., *New Astron.,* **12** (2006) 60; *New Astron.,* **12** (2007) 657.
[7] LEUBNER M. P., *Astrophys. J.*, **632** (2005) L1.
[8] KRONBERGER T., LEUBNER M. P. and KAMPEN E., *Astron. Astrophys.,* **453** (2006) 21.
[9] CARDONE V. F., LEUBNER M. P. and DEL POPOLO A., *Mon. Not. R. Astron. Soc.*, **414** (2011) 2265.
[10] CARVALHO J. C., NASCIMENTO Jr. J. D., SILVA R. and MEDEIROS J. R., *Astrophys. J.,* **696** (2009) L48.
[11] BETZLER A. S. and BORGES E. P., *Astron. Astrophys.,* **539** (2012) A158.
[12] DE FREITAS D. B. DE and MEDEIROS J. R. DE, *EPL*, **97** (2012) 19001.
[13] SILVA J. R. P., NEPOMUCENO M. M. F., SOARES B. B. and DE FREITAS D. B., *Astrophys. J.,* **777** (2013) 20.
[14] MARTINS, C. F. LIMA J. A. S. and CHIMENTI P., *Mon. Not. R. Astron. Soc.*, **449** (2015) 3645.
[15] LIMA J. A. S., SILVA R. and SANTOS J., *Phys. Rev. E*, **61** (2000) 3260.
[16] LIU L. Y. and DU J. L., *Physica A*, **387** (2008) 4821.
[17] LIU Z. P., LIU L. Y. and DU J. L., *Phys. Plasmas*, **16** (2009) 072111.
[18] GONG J. Y., LIU Z. P. and DU J. L., *Phys. Plasmas*, **19** (2012) 083706.
[19] LEUBNER M. P. and VOROS Z., *Astrophys. J.*, **618** (2005) 547.
[20] DU J. L. and SONG Y. L., "Solar wind speed theory and the nonextensivity of solar corona," in *Proceedings of the Third UN/ESA/NASA Workshop on the International Heliophysical Year 2007 and Basic Space Science*, *Astrophysics and Space Science Proceedings*, edited by H. J. Haubold and A. M. Mathai, pp. 93 (Springer, Berlin, 2010).
[21] BURLAGA L. F. and VINAS A. F., *Physica A* **356** (2005) 375.
[22] LEUBNER M. P., *Astrophys. J.*, **604** (2004) 469; *Astrophys. Space Sci.,* **282** (2002) 573.

[23] DU J. L., *Phys. Plasmas*, **20** (2013) 092901 and the references therein.
[24] BORGOHAIN D.R., SAHARIA K., GOSWAMI K.S., *Phys. Plasmas* **23** (2016) 122113.
[25] BACHA M., GOUGAM L.A., TRIBECHE M., *Physica A* **466** (2017) 199.
[26] BENTABET K., MAYOUT S., TRIBECHE M., *Physica A* **466** (2017) 492.
[27] LIVADIOTIS G., MCCOMAS J. D., *J. Geophys. Res.,* **114** (2009) A11105.
[28] LIVADIOTIS G., MCCOMAS J. D., *Space Sci. Rev.,* **175** (2013) 183
[29] LIVADIOTIS G., *J. Geophys. Res. Space Phys.,* **120** (2015) 880.
[30] LIVADIOTIS G., *J. Geophys. Res. Space Phys.,* **120** (2015) 1607.
[31] LIVADIOTIS G., *Entropy,* **17** (2015) 2062.
[32] VASYLIUNAS V. M., *J. Gerophys. Res.* **73** (1968) 2839.
[33] CRANMER S. R., *Astrophys. J.*, **508** (1998) 925.
[34] SILVA R. JR., PLASTINO A. R. and LIMA J. A. S., *Phys. Lett. A*, **249** (1998) 401.
[35] LIMA J. A. S., SILVA R. and PLASTINO A. R., *Phys. Rev. Lett.*, **86** (2001) 2938.
[36] DU J. L., *Chin. Phys. B,* **19** (2010) 070501, and arXiv:1012.2765.
[37] DU J. L., *J. Stat. Mech.*, (2012) P02006.
[38] GUO R. and DU J. L., *J. Stat. Mech.*, (2013) P02015.
[39] GUO R. and DU J. L., *Ann. Phys.*, **359** (2015) 187.
[40] DU J. L., *Europhys. Lett.*, **67** (2004) 893.
[41] DU J. L., *Astrophys. Space Sci.*, **312** (2007) 47.
[42] DU J. L., *Phys. Lett. A*, **329** (2004) 262.
[43] YU H. N. and DU J. L., *Ann. Phys.,* **350** (2014) 302.
[44] DU J. L., *Europhys. Lett.* **75** (2006) 861.
[45] ZHENG Y. H., LUO W., LI Q. and LI J., *EPL*, **102** (2013) 10007.
[46] ZHENG Y. H., *EPL*, **101** (2013) 29002.
[47] ZHENG Y. H., *EPL*, **102** (2013) 10009.
[48] ZHENG Y. H., and DU J. L., *EPL*, **105** (2014) 54002.
[49] ZHENG Y. H., and DU J. L., *Physica A*, **420** (2015) 41.
[50] GONG J. Y. and DU J. L., *Phys. Plasmas,* **19** (2012) 063703.
[51] LIU Z. P. and DU J. L., *Phys. Plasmas,* **16** (2009) 123707.
[52] VOTYAKOV E. V., MARTINO A. and GROSS D. H. E., *Eur. Phys. J. B*, **29** (2002) 593.
[53] KREMER G. M. *Interaction of Mechanics and Mathematics: Introduction to the Boltzmann Equation and Transport Processes in Gases* (Springer) 2010.
[54] KITCHATINOV L. L., arXiv:1108.1604
[55] DE FREITAS D. B. DE, NEPOMUCENO M. M. F., SOARES B. B. and SILVA J. R. P., *EPL*, **108** (2014) 39001.
[56] FIX J. D. *Astronomy: Journey to the Cosmic Frontier, fourth edition.* (McGraw-Hill Higher Education) 2006.